\documentclass[format=acmsmall, review=false]{acmart}
\usepackage{acm-ec-24-proc}
\usepackage{booktabs} 
\usepackage[ruled]{algorithm2e} 
\usepackage{natbib} 

\SetAlFnt{\small}
\SetAlCapFnt{\small}
\SetAlCapNameFnt{\small}
\SetAlCapHSkip{0pt}
\IncMargin{-\parindent}

\setcitestyle{authoryear}

\usepackage{dsfont}
\usepackage{siunitx}
\usepackage{listings}
\usepackage{xcolor} 
\usepackage{hyperref}

\ccsdesc[300]{Applied computing~Economics}
\ccsdesc[500]{Theory of computation~Exact and approximate computation of equilibria}
\ccsdesc[500]{Theory of computation~Multi-agent reinforcement learning}

\keywords{mechanism design, approximate equilibria, multi-agent reinforcement learning}

\acmYear{2024}\copyrightyear{2024}
\setcopyright{acmlicensed}
\acmConference[EC '24]{Conference on Economics and Computation}{July 8--11, 2024}{New Haven, CT, USA}
\acmBooktitle{Conference on Economics and Computation (EC '24), July 8--11, 2024, New Haven, CT, USA}
\acmDOI{10.1145/3670865.3673644}
\acmISBN{979-8-4007-0704-9/24/07}

\title[Understanding Iterative Combinatorial Auction Designs via MARL]{Understanding Iterative Combinatorial Auction Designs via Multi-Agent Reinforcement Learning}

\author{Greg d'Eon}
\affiliation{%
    \institution{University of British Columbia} 
    \city{}
    \country{Canada}}
\author{Neil Newman}
    \affiliation{%
    \institution{University of British Columbia} 
    \city{}    
    \country{Canada}}
\author{Kevin Leyton-Brown}
\affiliation{%
    \institution{University of British Columbia} 
    \city{}
    \country{Canada}}

\begin{abstract}
Iterative combinatorial auctions are widely used in high stakes settings such as spectrum auctions. 
Such auctions can be hard to analyze, making it difficult for bidders to determine how to behave and for designers to optimize auction rules to ensure desirable outcomes such as high revenue or welfare.
In this paper, we investigate whether multi-agent reinforcement learning (MARL) algorithms can be used to understand iterative combinatorial auctions, given that these algorithms have recently shown empirical success in several other domains.
We find that MARL can indeed benefit auction analysis, but that deploying it effectively is nontrivial. We begin by describing modelling decisions that keep the resulting game tractable without sacrificing important features such as imperfect information or asymmetry between bidders.
We also discuss how to navigate pitfalls of various MARL algorithms, how to overcome challenges in verifying convergence, and how to generate and interpret multiple equilibria.
We illustrate the promise of our resulting approach by using it to evaluate a specific rule change to a clock auction, finding substantially different auction outcomes due to complex changes in bidders' behavior.
\end{abstract}

\begin{document}

\maketitle

\section{Introduction}\label{sec:rl_intro}
Iterative combinatorial auctions are a high-stakes, socially important class of mechanisms.
These auctions have become the dominant mechanism for radio spectrum privatization \cite{milgromawards}, a domain we will use as a running example throughout this paper. Spectrum auctions revenues frequently reach billions of dollars and outcomes have national impact.
Iterative combinatorial auctions are also deployed in a wide range of other settings~\cite[see, e.g.,][]{PalaciosHuerta2021CombinatorialAI}.

It is important for bidders to make good strategic choices in an auction whose outcome can materially affect their company's value. 
It is also important for designers to establish auction rules that achieve good outcomes in terms of revenue, welfare, and other metrics.
However, neither task is easy, because iterative combinatorial auctions are profoundly complex. Combinatorial designs explode bidders' action spaces, but are essential when bidders have strong needs to express complementarities and/or substitutabilities across individual goods. 
Iterative designs yield a second exponential increase in strategic complexity, but are critical when bidders need to perform ``price discovery'' by updating their stated preferences in response to information about others' demand. 

Consider the case of spectrum auctions. 
The iterative combinatorial auctions used to sell radio spectrum can have 50+ page manuals explaining how they evolve over time and which bids are legal to submit \cite{appendixD}. 
They are held too infrequently to learn about from historical data, occurring only once every few years nationally. 
What's more, while there are a few popular auction formats, such as Simultaneous Multiple Round Auctions (SMRA) \cite{puttingauctiontheory}, Combinatorial Clock Auctions (CCA) \cite{cca}, and Clock Auctions\footnote{Clock Auctions are combinatorial auctions, but are a distinct auction format from Combinatorial Clock Auctions.} \cite{10.1257/0002828043052330}, even within a single format, rules tend to differ from one auction to the next.
For example, between the Canadian 700 MHz \cite{ised700} and 600 MHz \cite{ised600} auctions, the activity rule was altered; later, between the Canadian 3500 MHz \cite{ised3500} and 3800 MHz \cite{ised3800} auctions, a spectrum cap was added, limiting the maximum a bidder could win in each region.
With rules in constant flux, it is important to develop methods for reasoning about the effects of counterfactual policy changes in these auctions.
Ideally, we would analyze such problems by finding the Bayes--Nash equilibria of the game induced by each set of auction rules and evaluating how outcomes change in equilibrium. This is the dominant approach taken by theoretical work in economics, and has also been adopted by various computational approaches \cite[e.g.][]{positronic, THOMPSON2017583, RABINOVICH2013106}.
However, there is little hope of identifying Bayes--Nash equilibria of realistic iterative combinatorial auctions, as they are too complex to admit pen-and-paper analysis using known techniques.
Thus, the most influential existing work studies highly simplified settings.
For instance, for one auction format, \citet{riedel2006immediate} proved that it is an equilibrium for bidders to end the auction in the first round, but their proof applies only to auctions with a single product, infinitesimal price increases, and complete information about bidders' values.
Combinatorial action spaces and multi-round structures imply enormous extensive-form representations, putting them out of reach of traditional equilibrium solvers as well. 

An alternative is to simulate the auction under the assumption that bidders play some given, fixed strategy \cite[e.g.,][]{incentive, Cary2007GreedyBS}. 
One can then observe both the effect of a rule change and its sensitivity to changes in bidders' valuations and to randomness in the auction mechanism.
However, combinatorial auctions are rarely strategyproof \cite[see, e.g.,][]{simplebidding, Levin2016PropertiesOT}.
There is thus typically not a singular reasonable strategy to simulate, let alone one with consistent incentive properties across the rules being studied.

Multi-agent reinforcement learning (MARL) offers a middle ground:
more computational tractability than classical Bayes--Nash equilibrium computation, but richer strategic reasoning than static simulations.
Over the past decade, MARL algorithms have made great strides, achieving superhuman performance in two-player zero-sum games, such as Go \citep{Silver2016MasteringTG}, Shogi \citep{Silver2017MasteringCA}, and Stratego \citep{Prolat2022MasteringTG}.
These games share the property that players do not need to solve the equilibrium selection problem~\citep{eqselection}: if a player plays their part of any equilibrium, they will do at least as well against any other opponent equilibrium strategy.
Of course, this property does not hold in auctions, nor in zero-sum games having more than two players. 
It is thus striking that recent work has shown superhuman performance in multiplayer zero-sum games such as poker \citep{Brown885}. 
Of particular relevance to what follows, Counterfactual Regret Minimization (CFR)~\citep{Zinkevich2007RegretMI} is a popular, modern MARL algorithm that was recently used to solve Texas Hold 'em poker~\cite{bowlingpoker, libratus}. 
Also impressive was the demonstration of human-level play in the board game Diplomacy~\citep{anthony2022learning, Bakhtin2022HumanlevelPI}, a game sharing other key similarities with our domain: a combinatorial action space and iterative, simultaneous moves.

We foresee no near-term interest from market participants in deploying ``superhuman'' bots to bid autonomously in complex and high-stakes auctions without human oversight---nobody is going to bet a Fortune 500 company's future on an RL algorithm running against unknown opponent strategies in a novel economic environment. 
However, even in such domains, MARL offers promise for evaluating competing mechanisms and gaining economic understanding.
Tools from deep learning and reinforcement learning have been used to 
find revenue-optimal designs for small, single-stage auctions~\cite{Dutting2019RegretNet, Dutting2023DiffEcon, Thang2020BanditAlgorithm},  
optimize reserve prices in repeated auctions~\cite{Tang2017ReinforcementMechanismDesign, Ai2022RLAuctionDesign},
and react to opponent policies in ad auctions~\cite{Jin2018RealTimeBiddingMARL,Banchio2022ArtificialIA}.
MARL algorithms have also been used to find equilibria in single-round combinatorial auctions~\cite{Bosshard2020BNEAuctions,bichler2021learning, bichler2023computing}
and multi-round dynamic or sequential mechanisms~\cite{Thoma2023PBE,Thoma2023Dynamic,Pieroth2023equilibrium}.
Closest to our precise domain, \citet{pacaud2023bidding} combine heuristic strategies with Monte-Carlo tree search to find effective bidding strategies in clock auctions; however, they focus on finding strategies that perform well against existing heuristics in auctions with complete information.
Beyond auctions, MARL has been used to identify and evaluate complex policy decisions regarding taxation~\cite{zheng2022ai}, study regulations on platform economies~\cite{parkesplatforms}, and study bargaining dynamics of learning algorithms~\citep{Abramowicz2020ModelingSB, Li2023CombiningTG}.







This paper aims to use MARL to gain economic insight in (incomplete-information) iterative combinatorial auctions. Such insights could help a human bidder assemble a strategic ``playbook'' by providing examples of strong bidding both by an agent with their own preferences and by their counterparties; to date, such playbooks are typically derived via pen-and-paper analysis of dramatically simplified subproblems and via observations of expert human play in a handful of ``mock auctions''. Insights gleaned from MARL analysis could also help an auction designer to trade off the costs and benefits of candidate rule changes, evaluating economic variables such as revenue, welfare, length of the auction, variance in outcomes, etc.
Unfortunately, actually running MARL to understand an iterative combinatorial auction is not nearly as simple as implementing an auction simulator and unleashing multiple copies of some off-the-shelf algorithm. 
Modeling a spectrum auction verbatim from its rulebook produces games that are vastly too large to solve.
One is then immediately faced with important modeling choices, finding ways to reduce the game to a feasible size without abstracting away the most strategically important elements.

Our work makes two main contributions. First, in Section~\ref{sec:rl_cma}, 
we present a general methodology for studying complex, iterative auctions using MARL. 
Notably, we recommend simplifying the auction model by abstracting the action space, controlling the number of bidders, and limiting the auction's length, without giving up on qualitatively important features such as imperfect information or asymmetry between bidders.
We describe the landscape of MARL algorithms and important problems that can arise, such as arbitrary behavior when multiple actions lead to the same outcome, and convergence to brittle equilibria that involve unrealistic levels of coordination. 
We also discuss key challenges in verifying an algorithm's convergence, and interpreting trained policies in the face of multiple equilibria.

Second, we present a case study showing that, despite the challenges just described, MARL can be used to conduct novel, economically meaningful analysis. 
We consider a prominent family of iterative combinatorial auctions called clock auctions and, leveraging two modern MARL algorithms, investigate a key design choice: how to process bids when multiple bidders want to reduce demand (submit a drop bid) for the same product to a level below supply.
Section~\ref{sec:rl_experiments} describes the problem, how we model it as a game, how we solved it using MARL, and how we assessed performance. Section~\ref{sec:results} then goes on to demonstrate that this rule change does indeed give bidders incentive to change their behavior, yielding substantial differences in both revenue and auction length.
Furthermore, we show that a designer would incorrectly conclude that the rule change has the \textit{opposite} effect if they instead modeled bidders as following a straightforward, myopic bidding heuristic.

Beyond these findings, we contribute a highly configurable clock auction environment. In the hopes that it will serve as a testbed for future research on MARL for auctions, we expose many parameters representing possible auction rule changes. 
We also provide software that generates clock-auction game instances and associated bidder values, leveraging a realistic value model from the literature~\cite{Weiss2017SATSAU}. 
All of our code is available at \url{https://github.com/newmanne/open_spiel}.
The paper concludes in Section~\ref{sec:rl_discussion}, where we discuss various design choices ripe for future investigation and speculate on methodological advancements.

\section[Methodology: Understanding Iterative Combinatorial Auctions via MARL]{Methodology: Understanding Iterative Combinatorial Auctions via Multi-Agent Reinforcement Learning}\label{sec:rl_cma}
We begin with our methodology for applying MARL to understand iterative combinatorial auctions. 
First, we discuss how we model an auction, trading off model fidelity with size of the resulting game tree. 
We then turn to ``solving'' the game using MARL, discussing the strengths and weaknesses of different MARL algorithm families, the problem of multiple equilibria, and equilibrium refinements (preferring pure equilibria; avoiding brittle equilibria). 
We conclude by considering how to validate and interpret policies: assessing the extent to which policies have converged to a Bayes--Nash equilibrium and reasoning across multiple equilibria.

\subsection{Modelling an Auction}
It is tempting to simply model an auction wholesale, mapping its entire rules manual into a MARL environment, and unleash a MARL algorithm on it.
Unfortunately, this approach is unlikely to succeed for iterative combinatorial auctions: modeling many bidders, bundles, and rounds will yield an enormous game tree that is simply infeasible to solve.
Much like pen-and-paper analysis, one must abstract away certain elements to buy the ability to accurately model others.

Certain kinds of rule complexity that would pose significant barriers to pen-and-paper analysis are unproblematic for MARL, such as winner determination algorithms that fall back on default behaviors when solvers time out, ideosyncratic tie-breaking rules, and complex activity rules determining conditions under which certain actions are available to certain bidders. 
In other cases, auction elements can dramatically increase the number of actions available to a player in each information set, such as opportunities for intra-round bidding, the option to sidestep activity rules a small number of times, or the ability to place jump bids.

We recommend gaining a solid foundation by starting with simple, easy-to-solve games and scaling up later rather than starting with complex ones that might prove to be intractable.
We have been surprised countless times by how even auctions with moderate numbers of bidders, products, and rounds can lead to enormous, unwieldy game trees. 
However, as we demonstrate in our experiments in Section~\ref{sec:rl_experiments}, there is a sweet spot of game sizes out of reach of pen-and-paper analyses, yet tractable for MARL.
Here, we describe several high-level decisions that a designer can make to control the size of the game tree.
The key metric of importance is usually the number of distinct \textit{information states}---the number of game states that an agent can distinguish.

\subsubsection{Number of actions.}\label{subsubsec:actions}
Combinatorial auctions typically have enormous action spaces, often corresponding to the set of all legal bundles.
The action space determines the branching factor of the game tree, so reining it in has a dramatic effect on game size.
One helpful way to limit the size of an action space is to experiment with auctions that have few products or units of each product available, limiting the number of possible bids.
However, we caution against reducing the auction down to a single product (e.g., several units of a generic license), as auctions with only one product may not engage with many of the auction's rules, leaving interesting behavior unstudied. 
We recommend beginning with auctions offering two or three products.

It is also possible to study auctions with more available bundles by restricting bids, only allowing bidders to choose from a small number of heuristic strategies at each state.
For example, these heuristics could include bidding myopically, maintaining the last round's bid, bidding on an opponent's holdings, or attempting to drop out of the auction.
Care is required to ensure that this restriction does not significantly hamper players' abilities to best respond to others' strategies. 
As a concrete example, it may be tempting to shrink the action space by limiting bidders to only bidding on packages they genuinely value. 
However, it is well-known that there are auction instances where such \textit{simple bidding} is never optimal in Bayes--Nash equilibrium~\citep{simplebidding}.
A helpful intermediate option is to allow players to deviate from a given set of heuristics up to $k$ times per auction; this can be useful when it is only strategically important to deviate from known heuristics occasionally, and also can be used to diagnose the effectiveness of the given set of heuristics.
This approach has the beneficial side effect of producing strategies that are easier for an analyst to interpret, which can be particularly helpful for developing a playbook. 
Restricting actions to a smaller set is also likely to make the resulting strategies more useful as meta-strategies in Empirical Game Theory Analysis \citep{wellman2006methods}.

In some auctions, a naive implementation would produce a continuous action space.
For instance, both intra-round bidding in clock auctions and the supplementary rounds of CCAs give bidders the ability to report real-valued prices.
Continuous action spaces pose a problem for many MARL algorithms,
which commonly assume that the action space is discrete.
Thus, when possible, we recommend to avoid modelling the part of the auction that admits a continuous action space.
When this is impractical, it may be possible to discretize the action space: 
for instance, allowing bidders to only bid in \$1 increments, or again employing a finite set of heuristics (each of which maps to an unrestricted real value).
Alternatively, one can use a MARL algorithm that is designed specifically for continuous action spaces; we caution that these algorithms can make strong assumptions about agents' policies, requiring bidders to make deterministic bids~\cite{Lowe2017MADDPG} or to draw their bids from a parametric (e.g., Gaussian) distribution.

\subsubsection{Auction length.}
Real-world spectrum auctions can last tens to hundreds of rounds. 
This is problematic for MARL analysis, as the size of a game tree typically grows exponentially with its maximum length: with $r$ rounds and $B$ possible bids each round, a game tree would have $B^r$ states!
However, there is often little value in analyzing such long auctions.
In practice, it is common to see the same bids repeated at slightly increasing prices for many rounds, with much of the action compressed into a few key rounds.
We recommend setting opening prices and price increments such that auctions last no longer than ten rounds; this tends to avoid enormous game trees without sacrificing much of the strategic complexity of the auction. 
In some cases, even two or three rounds can be enough to yield economically meaningful conclusions.

A related technical issue is that, in theory, auctions can be arbitrarily long: as long as bidders continue to overdemand products, the auction will continue.
These infinitely large game trees do not typically pose a problem to theorists, who can handle them with inductive proofs, nor to reasonable bidders in practice, who would never expose themselves to indefinitely rising prices.
However, they \textit{do} pose a problem to most RL algorithms, which cannot learn to avoid each of these unreasonable bid trajectories without first gaining experience by making them.
We recommend constraining agents' bids to rule out such arbitrary-length bidding sequences.
A natural solution is to disallow bidders from making bids that are strictly dominated by dropping out of the auction, ensuring that they are forced to drop out when prices become sufficiently high.
Another is modelling bidders as budget-constrained, disallowing bids that would exceed a bidder's budget if processed.
Budgets are a more tenuous modelling choice: while real-world bidders do have finite resources, it may be unrealistic to prohibit a bidder from buying a license that they value more than its cost.
We recommend against adding an artificial limit on the length of the auction.\footnote{
We experimented with directly limiting the auction's length, giving all players arbitrary, negative utilities upon hitting this limit. We recommend against this as it can have a dramatic effect on the equilibria of the game. Specifically, we found that weak bidders that would otherwise be allocated nothing could make non-credible threats to bid up to the limit, forcing stronger bidders to concede an item.}

\subsubsection{Number of bidders.}
In practice, spectrum auctions can involve dozens of bidders.
However, many bidders represent small, locally-operating telecoms that have limited ability to make large-scale bids.
If smaller bidders must be modeled,
consider whether a deterministic heuristic strategy (as opposed to a learned policy) suffices,
or whether their action spaces can be restricted to their bundles of interest. 

\subsubsection{Perfect or imperfect information.}
In a perfect-information game, each bidder knows their opponents' exact valuations. 
Perfect-information models of auctions often have equilibria in which bidders coordinate strongly and unrealistically. 
For example, in the perfect-information model of a sealed-bid, first-price auction, the bidder with the highest value bids the second-highest bidder's value and all other bidders bid arbitrarily; missing much of the strategic reasoning that takes place in such an auction. 
Analogously, when we experimented with perfect-information models of iterative combinatorial auctions, we often found that they terminated in a single round. 

Real auctions, where bidders likely have priors over their opponents' strategic goals and valuations, are best modeled as Bayesian games.
While Bayesian modeling is a big step mathematically, in an RL environment it amounts simply to adding a chance node at the root of the game tree. 
Such a node creates a copy of the game tree for each type combination, which leads to a much larger game tree but, since bidders do not know each others' types, has a smaller effect on the number of information states. 
We thus recommend experimenting with bidders that have a small number of types.
For example, when experimented with imperfect-information settings in which each bidder can have one of multiple possible valuations, we have typically observed more interesting, multi-round behavior, than in the perfect-information case, including pooling equilibria in which weaker types act like their stronger counterparts to drive better negotiations than they ``deserve''.

\subsubsection{Asymmetry between bidders.}
It is common in auction analysis to model bidders as symmetric. This offers several technical advantages.
When all bidders are symmetric, the equilibrium selection problem can mitigated by searching for symmetric equilibria. 
Furthermore, if all agents behave symmetrically, only a single policy needs to be trained.
However, in reality, bidders in spectrum auctions are not usually well-approximated as symmetric, as they often have vastly different resources, business interests, and priorities.
Thankfully, such asymmetry is a straightforward addition to a MARL environment, as RL algorithms have little trouble dealing with asymmetric bidders.
We advocate for experimenting with bidders of varying relative strengths.

\subsection{Finding Equilibria}
We now turn to the problem of using MARL to find near-equilibrium strategy profiles.

\subsubsection{Choosing an algorithm.}
Multi-agent reinforcement learning is a broad field with a wide landscape of methods.
Here, we describe just two key dimensions along which MARL algorithms vary.
For a comprehensive overview of the field, see, e.g.,~\citet{Albrecht2024MarlBook}.

The first dimension is whether agents' policies have a tabular or function approximation representation.
In tabular methods, each information state in the game is represented separately; bidders faced with a new information state do not reason about its similarity to other states with which they have previous experience, instead learning afresh.
This is a double-edged sword: it does not constrain players' behavior in any particular way, allowing them to learn flexible strategies, but causes erratic behavior in rarely-encountered states.
With function approximation (i.e., deep reinforcement learning), each information state is represented by a feature vector and the policy is represented by a parametric function, such as a deep neural network.
This allows agents to generalize from one part of the game tree to another, but can be more difficult to train reliably.

A second key dimension distinguishing MARL algorithms is whether they explore counterfactual actions.
Many algorithms from single-agent reinforcement learning, such as policy-gradient methods~\cite{policygrad}, sample a single path through the game tree at each iteration, only learning about their rewards along that path.
While it is possible to use these algorithms in multi-agent settings by running them independently for each player, they can struggle due to focusing most of their training effort on a small fraction of the game tree.
At the other extreme, algorithms such as Counterfactual Regret Minimization (CFR)~\citep{Zinkevich2007RegretMI} enumerate every information state in each iteration.
This extra coverage allows these algorithms to compare rewards across different actions, but their runtime quickly becomes infeasible for larger games.
Algorithms can fit between these two extremes: for example, external-sampling Monte-Carlo CFR (MCCFR)~\citep{mccfr} enumerates one player's actions but samples opponent actions and chance outcomes. 

For large games, it can be infeasible to store a tabular policy. 
However, for smaller games, we see little reason not to use a tabular method that explores counterfactual actions, such as CFR; at the very least, we recommend such algorithms as a baseline for comparing to other approaches.

\subsubsection{Multiple equilibria.}\label{subsubsec:multipleq}
Complex auctions typically admit multiple equilibria.
When using non-deterministic MARL algorithms, a simple way to find multiple equilibria is to run the algorithm with multiple random seeds, which influence both its random initialization and any random events within the environment, causing it to take a different learning trajectory on each run.
We thus recommend non-deterministic algorithms, e.g., preferring Monte-Carlo sampling variants of CFR over the vanilla algorithm.
We speculate about potential approaches to identifying multiple equilibria with deterministic algorithms in Section~\ref{subsec:quality_diversity}.

When two or more actions lead to the same terminal reward, RL algorithms have no reason to prefer one action over another, and equilibria multiply.
This can be detrimental to both convergence and auction outcomes. 
For example, consider a bidder facing an incredibly strong opponent that will price them out of the auction no matter how they behave in early rounds.
Since they cannot impact their own utility, they can play any strategy in equilibrium, ranging from immediately dropping out of the auction to raising their opponent's prices up to their own value. Their arbitrary choice will impact many auction outcomes of interest (e.g., auction length, revenue).
We therefore advocate for adding small, ``secondary'' rewards to the RL environment, biasing agents toward certain strategies (e.g., preferring to drop out than to stay in) when all else is equal.

\subsubsection{Pure or mixed equilibria.}
An analyst must decide whether to focus solely on pure equilibria or to also consider mixed equilibria.
There are several advantages to studying pure equilibria.
First, working with pure equilibria is computationally beneficial: restricting to pure strategies greatly reduces the strategy space, tending to make it easier to train, validate, and evaluate policies.
Second, some MARL algorithms may not be able to converge to mixed equilibria at all.
For a broad class of MARL algorithms, the only stable points of the learning dynamics are pure equilibria \citep{donotmix}.
While there is no guarantee that a game will have any pure equilibria, in our experiments, we were able to find a pure equilibrium in the vast majority of our games.

We recommend considering whether it is sufficient to study pure equilibria.
If it is critical to allow bidders to play mixed strategies (e.g., to allow randomized bluffing), one must be cautious about the choice of MARL algorithm.
Otherwise, given that the outcome of a MARL algorithm is typically mixed, we recommend rounding learned policies to the nearest pure strategy, deterministically playing the action with the highest probability; we refer to such rounded policies as \textit{modal.}
This rounding step is important for MARL algorithms that play actions with probabilities that approach zero, but do so slowly, potentially never reaching zero.
One such example is PPO, which rewards entropy in its loss function.

\subsubsection{Avoiding brittle equilibria.}\label{subsec:findingeq}
Prior work in MARL has found that trained policies can fail catastrophically when facing new opponents.
For example, in the cooperative card game Hanabi~\citep{BARD2020103216}, policies trained through self-play can achieve near-perfect scores, but fail to get even a single point when playing with policies from other training runs.
One way this can happen is when MARL algorithms identify \textit{brittle} equilibria, in which players rely on perfect coordination with each other, which they achieve through their training history.
Such equilibria may have no robustness to even minor misspecifications. 
Furthermore, in the domain of spectrum auctions, achieving such coordination would likely require illegal collusion between the bidders.
To combat this problem, during training we recommend having policies \textit{tremble} during training, deviating from the policy and playing an action uniformly at random with a small probability (e.g., 1\%).
Such trembling pushes agents away from equilibria that demand perfect coordination.
Additionally, it serves as a way of further breaking down indifferences: even if two actions give the same rewards on-path, they may give different rewards off-path.

\subsection{Validating and Interpreting Policies}
Lastly, we discuss how to proceed after training a set of policies with a MARL algorithm.

\subsubsection{Assessing convergence.}\label{subsub:assess}
It is important to understand whether a MARL algorithm has converged to an equilibrium: if not, it is difficult to know whether trends in bidders' strategies are simply caused by poor convergence.
The ideal measure of convergence is the \emph{NashConv} of the policy, which is defined as the sum of each agent's \emph{regret}---the utility gain they would achieve by playing a best-response, holding their opponents' strategies fixed. 
A NashConv of zero thus corresponds to a Nash Equilibrium, 
and a positive NashConv corresponds to an $\epsilon$-Nash equilibrium, where $\epsilon$ is no greater than the NashConv.
Unfortunately, NashConv is nontrivial to compute, as it requires requires finding the optimal solution to a single-agent RL problem for each player, a task that requires an exhaustive search.
Thus, while it would be ideal to checkpoint the trained policy periodically and to analyze only the checkpoint with the smallest NashConv (which need not be the latest checkpoint), this is infeasible on all but the smallest of games.

When it is not possible to compute NashConv exactly, it can still be valuable to estimate it by computing approximate best responses.
One way to do so is to restrict the action space of the game to a smaller set of heuristics (such as in Section~\ref{subsubsec:actions}) and compute best responses on this restricted game.
Another is to run a traditional single-agent RL algorithm, which may find a good policy even if it has no guarantee of finding the optimal policy.
These approximate best responses give a lower bound on a player's regret, as the true best response must increase utility by at least as much as the found strategy.
While this approach cannot guarantee that a strategy profile is an equilibrium (because it can fail to find optimal best responses), it can still be useful to have some evidence that each player's strategy is not easy to improve upon.

Note that it is dramatically easier to compute NashConv of a pure strategy profile, as this exponentially reduces the number of subtrees that must be examined.
We recommend attempting to compute NashConv exactly if analyzing pure strategies. 
When one expects a pure strategy Nash equilibrium, it can also be useful to track the entropy of players' policies, which must decay to 0.

\subsubsection{Interpreting multiple equilibria.}\label{subsucsec:multieq}
We have already stated some refinements above that rule out certain equilibria. 
If a single equilibrium remains for each game, one can simply study its strategy profile or check how its outcomes are affected by changes to the auction.
However, the analysis is murkier if multiple equilibria remain: should an analyst treat one equilibrium as more important than another? Should the frequency with which a given equilibrium is observed have any bearing on its importance?
With no other desiderata claiming that one equilibrium is more plausible than another, we believe an analyst should not rank equilibria by their frequency, as the exact choice of MARL algorithm biases which equilibria are likely to be found. 
Thus, we recommend looking at the \textit{range} of possible strategies or outcomes represented by the set of equilibria, rather than looking at e.g., the mean outcome across all samples, or only studying a single sample. 

\section{Case Study: Bid Processing in Clock Auctions}\label{sec:rl_experiments}
We now demonstrate our methodology on clock auctions, a modern iterative combinatorial auction format. 
As a case study, we focus on a specific question about the design of this auction: in which order should the auctioneer process simultaneous bids?
We discuss how to instantiate each part of our methodology to tackle this question. 

\subsection{Clock Auctions}\label{sec:rl_clock}
Clock auctions are an iterative combinatorial auction format that have been used to allocate spectrum, e.g., in the FCC's Auction 102 \cite{fcc102} and the Canadian 3500 MHz auction \cite{ised3500}.
Here, we describe a simplified auction format that draws heavily upon these real spectrum auctions; we provide a formal description in Appendix~\ref{sec:clockauctions}.

In a clock auction for allocating spectrum, an auctioneer sells indivisible radio licenses to a number of bidders.
The licenses are divided into products, each corresponding to an equivalence class of geography and spectrum quality; there are typically several units of supply available for each product.
The auctioneer sets an opening price for each product.
Then, the auction proceeds over a series of rounds. 
In each round, bidders submit a vector-valued bid representing the number of units of each product they want to buy at the current prices. 
If any product is overdemanded---that is, has more aggregate demand than supply---then the auctioneer increases the price on each overdemanded product, informs all bidders of every product's aggregate demand and new price, and proceeds to the next round of the auction.
Otherwise, the auction ends, with each bidder winning the units they bid for, paying their final prices.

We highlight two additional rules that add considerable complexity to clock auction analysis. 
First, to discourage strategic bidding, bidders cannot bid for arbitrary bundles in each round. 
Future bids are constrained by past bids through an \textit{activity} rule.
We consider a simple implementation which assigns each bundle a number of ``eligibility points'' and requires bids to be weakly decreasing in eligibility points.
Second, to avoid leaving licenses unsold, the auctioneer rejects bids that would lower the demand for a product below its supply. 
Combined with the activity rule, this creates a potential exposure problem, as bids may only be partially processed if points from a dropped product are needed to pick up a new product.

We now use our methodology to evaluate a potential design choice: \textit{when multiple bidders simultaneously request to drop licenses, how should the auctioneer determine which bids to process first?}
Processing every bidder's drop request could change a product from being over- to under-demanded, which the auction rules disallow.
To avoid leaving licenses unsold, the auctioneer must instead process the drop requests sequentially. 
In practice (e.g., in the Canadian 3500 MHz auction), these ties are handled by randomly ordering the queue of requests and processing entire drop requests in this order, letting bidders drop as many licenses as they requested, until no excess demand remains.
This ``drop-by-bidder'' mechanism can have high variance, processing either all or none of each bidder's submitted bid.
We compare this mechanism to an alternative where a request to drop several licenses of a product is represented in the queue as multiple requests to drop an \textit{individual license}.
This alternative ``drop-by-license'' mechanism reduces variance, making it possible for each bidder to have some of their drop requests processed.
These bid processing mechanisms are well-suited for our computational methodology because they require a large amount of case-based reasoning to study, making them overwhelming to analyze with traditional methods but ideal for reinforcement learning.

\subsection{Defining the Environment}\label{subsec:rl_def_env}
In the remainder of this section, we describe how we instantiated the methodology to compare these two auction designs.
We implemented our clock auction game in the framework of OpenSpiel \citep{LanctotEtAl2019OpenSpiel}.

\subsubsection{Auction setup.} 
We consider auctions with two bidders and two products: an unencumbered\footnote{An \textit{encumbered} license only covers a fraction of the bandwidth that a full, unencumbered license would. Encumbered licenses are usually priced at a discount to reflect their poorer coverage.} 
product with one license of supply, and an encumbered product with four licenses, each of which only gives 60\% of the bandwidth of an unencumbered license.
Each license has eligibility points and an opening price proportional to its bandwidth.
We focus on a specific state midway through the auction\footnote{We do not assume this particular midway point is part of any equilibrium of the game starting from the first round. We are interested in equilibrium behavior starting from this point.}: we assume that each bidder has previously bid for three of the encumbered licenses for several rounds, making it more expensive per MHz of spectrum than the unencumbered license.
The bidders have a strategic choice to make: continue bidding on the more expensive product, or attempt to switch to the cheaper one (e.g., dropping two encumbered licenses to pick up one unencumbered license). 
The bid processing mechanism is triggered if both bidders attempt to switch.
In the drop-by-bidder mechanism, the auctioneer chooses a random order over the bidders and processes all of one bidders' drops in order, allowing one bidder to switch while the other's bid is unprocessed.
In the drop-by-license mechanism, the auctioneer instead processes the individual requested drops in a random order, most often ending with both bidders dropping one encumbered license. In this case, the activity rule prevents both bidders from picking up the unencumbered license.\footnote{Real auctions often include a \textit{grace period} rule, giving bidders a chance to regain lost activity in the following round. We do not model a grace period. However, notice that in this strategic setting, if both players lose activity, the auction is guaranteed to end, and a grace period rule would have no effect in these cases.}

We set the opening start-of-round prices to \$12 million on the unencumbered product and \$7 million on the encumbered product.
We set the clock speed---the price increase after each round---to 5\% of the current price.
Thus, after two rounds of bidding for the encumbered product, its price is \$7.72 million.

\subsubsection{Bidder utilities.}
We assume that each bidder has a finite number of equally likely types.
Each type has a quasilinear utility function, meaning their utility for winning a bundle is its value minus its price paid. 
We choose these values by drawing them from the MRVM value model~\cite{Weiss2017SATSAU}, informed by the 2014 Canadian spectrum auction.
MRVM models bidders as having a critical amount of spectrum which they need in order to provide a high quality of service, and low marginal value when they have much more or less than this critical amount.
Bidders are thus described by two parameters: their \textit{value per subscriber}, the amount of value they would gain from winning all of the spectrum, and their \textit{market share}, the fraction of spectrum which they need to obtain half of this maximum value.
Their values are then sigmoidal with respect to the fraction of spectrum they win, with high marginal values within $\pm$15\% of their market share, and low marginal values outside of this range.\footnote{We note a similar sigmoidal modeling choice in \citep{Bichler2022TamingTC}, which studies the FCC's C-band.}
For each type, we draw their value per subscriber uniformly at random between \$20 million and \$30 million, and their market share uniformly at random between 35\% and 50\%.
Note that these non-linear value functions can make ``switch'' bids risky, as bidders can lose a large amount of value if a switch bid is only partially processed.

In preliminary experiments, we found that some samples from this value model led to games that were either strategically uninteresting (e.g., had a bidder too weak to have a chance of winning a single license) or had game trees too large to be feasible (e.g., could last 20 rounds or more). 
To guard against these problems, we ran rejection sampling, resampling values that exhibited one of these problems.
We generated values for bidders having 1, 2, 3, 5, and 7 types, generating 5 value profiles for each number of types.
We provide the precise details of the rejection sampling process and an example of bidders' value functions in Appendix~\ref{sec:sats_details}.

\subsubsection{Secondary rewards.}
With no further changes, different actions can lead to the exact same terminal rewards: 
for example, when a product's demand is already at supply, attempting to drop one license often has the same effect as dropping all of the licenses.
We address this problem by adding secondary rewards (as described in Section~\ref{subsubsec:multipleq}).
Specifically, each round, we subtract a penalty between zero and one from each player's utility.
A penalty of zero corresponds to choosing the most profitable bundle at the current prices, and a penalty of one to the least profitable; other bundles interpolate between these two extremes.
These secondary rewards can be understood as giving bidders an incentive to prefer straightforward bids and shorter auctions all else being equal.
These penalties are removed after training, and all of our reported NashConv values are with respect to the unmodified game.

\subsection{Finding Equilibria}
We ran two MARL algorithms: a Monte Carlo sampling variant of Counterfactual Regret Minimization, and Proximal Policy Optimization, an on-policy deep reinforcement learning method.

\subsubsection{Monte Carlo Counterfactual Regret Minimization}
Counterfactual Regret Minimization (CFR) is a tabular MARL algorithm.
It is effectively a generalization of regret matching to imperfect information extensive form games: at each information state, it tracks a \textit{regret} for each action, playing actions with probabilities proportional to their regret.
Specifically, we use a variant known as External Sampling Monte-Carlo CFR~\cite{mccfr}, which avoids traversing the entire game tree by randomly sampling chance outcomes and opponent actions.
Initial experimentation revealed that two common performance tricks were helpful: we use the regret matching plus~\cite{cfrplus} algorithm instead of regret matching, and we use linear CFR~\cite{linearcfr}, which discounts regrets over time so that later iterations have larger impacts.
Lastly, as is common for CFR, we analyze the average strategy across all iterations, rather than the last iterate.

We make one modification to this standard algorithm.
To help steer MCCFR away from brittle equilibria, we train against \textit{trembling} opponents: 
when we sample an opponent's action, we pick an action uniformly at random 1\% of the time, and otherwise sample from their trained policy.
This trembling only occurs during training; there is no trembling in our final policies. 

\subsubsection{Proximal Policy Optimization}
\label{sec:ppo}
Our second algorithm is a deep reinforcement learning algorithm, Proximal Policy Optimization (PPO) \citep{DBLP:journals/corr/SchulmanWDRK17}. 
PPO is an on-policy algorithm inspired by classic policy gradient methods.
It represents its policy with a neural network. Roughly, to train this policy, it collects a sample of information states, actions, and their eventual rewards, then updates its policy to increase the probability on actions leading to high rewards.
PPO differs from a standard policy gradient method in that its objective function penalizes large changes in action probability, causing it to change its policy more slowly during training.

\begin{figure}
    \centering
    \includegraphics[width=10cm]{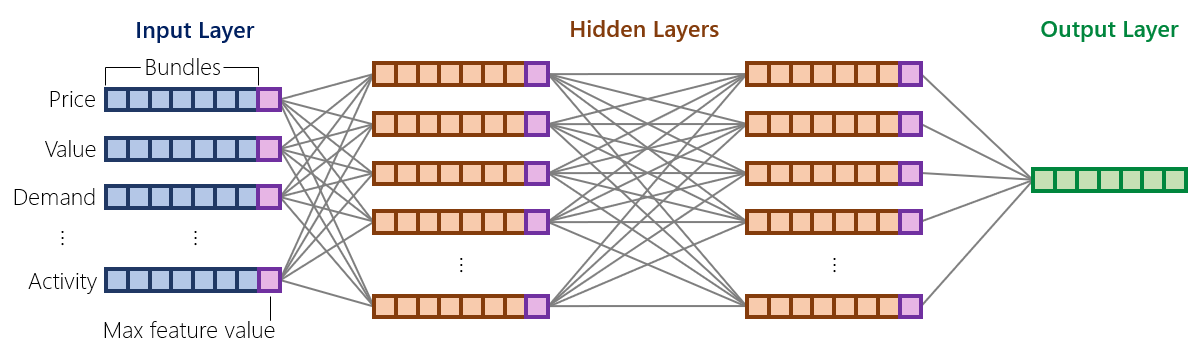}\hfill
    \caption{The AuctionNet neural network architecture.}
    \label{fig:network}
\end{figure}

In order to use PPO, we must choose a neural network architecture.
We experiment with two.
The first is a feed-forward neural network, accepting a flat vector describing all variables known in the information state as input.
While this is a simple architecture, it is unlikely to be effective: such a network would have to learn how to process facts about every single bundle of licenses independently.
To address this issue, we also test an equivariant architecture~\cite{Hartford2018DeepMO}, which is shown in Figure~\ref{fig:network}.
In this architecture, each bundle has an associated set of features, and the network must perform the same computation on every bundle. 
Comparisons can be made across bundles using max-pooling. 
This enforces that the network must apply the same function to each bundle, giving it a strong inductive bias toward learning a single, simple function and applying it globally.
We list the variables given to these neural networks in Appendix~\ref{sec:ppo_features}.

For both MCCFR and PPO, we convert all trained policies to pure strategies, placing probability 1 on the modal action.
This conversion is computationally important, making it possible to compute NashConv as described in Section~\ref{subsub:assess}.

\subsection{Validating and Interpreting Policies}
We compute several metrics about each of our trained policies.
For each simulation, we record the revenue, welfare, length (in rounds), number of unsold licenses, and number of lotteries in bid processing.
The latter is relevant to the design decision being studied, demonstrating whether agents are engaging in strategies that trigger lotteries or avoiding them.
To validate convergence, we compute NashConv with a standard depth-first search algorithm.

We also compare our results to simulations of \textit{straightforward bidders}: bidders who,
in each round, bid for the bundle with the highest utility at the current prices.
This is a natural definition of straightforward bidding behavior as it is myopic, with no regard to price increases or activity constraints in future rounds.\footnote{In particular, this definition of straightforward bidding is ignorant of the fact that any bid to lower demand could be rejected in part or in full by the bid processing algorithm, leading to exposure. It would be interesting to extend this heuristic to incorporate a demand forecasting model, predicting whether a drop bid is likely to be processed.}
It is also analogous to existing definitions of straightforward bidding in simultaneous multi-round auctions~\cite[e.g.,][]{milgrom2004putting}, a related auction format.

\section{Experimental Results}\label{sec:results}
We now discuss three sets of experiments and their results.
First, we tuned our two MARL algorithms. We found configurations that could identify policies with low (often zero!) NashConv in our games.
Second, we ran our tuned algorithms on games with two different bid processing algorithms. We observed that the modification to bid processing yields substantially different auction outcomes due to non-trivial changes in bidder behavior.
Third, we tested how well these algorithms scaled to larger games. We found that they often still converged to approximate equilibria.

Our computational environment is described in Appendix~\ref{sec:computationalenv}.
Except where noted otherwise, we ran our MARL algorithms for one hour of walltime,
computing NashConv once at the end of the run.

\subsection{Ablations and Hyperparameter Tuning}
First, we experiment with various hyperparameter settings for each of our MARL algorithms.

\subsubsection{MCCFR: Trembling Opponents and Secondary Rewards}
MCCFR has few hyperparameters that need to be tuned.
Here, we evaluate our recommendations from Sections~\ref{subsubsec:multipleq} and \ref{subsec:findingeq}, where we advocated for breaking ties between rewards for actions, and for having opponents tremble to avoid brittle strategies. 
To test these techniques, we compared four variants of MCCFR, varying whether opponents trembled during training and whether the game included secondary rewards.
We ran each variant with 10 random seeds on 10 different 5-type games (comprising of 5 valuation samples and 2 bid processing rules): a total of 100 runs for each variant.
For each run, we computed NashConv at six checkpoints. 
Runs took between 10 and 60 minutes.

\begin{figure}
    \centering
    \includegraphics[]{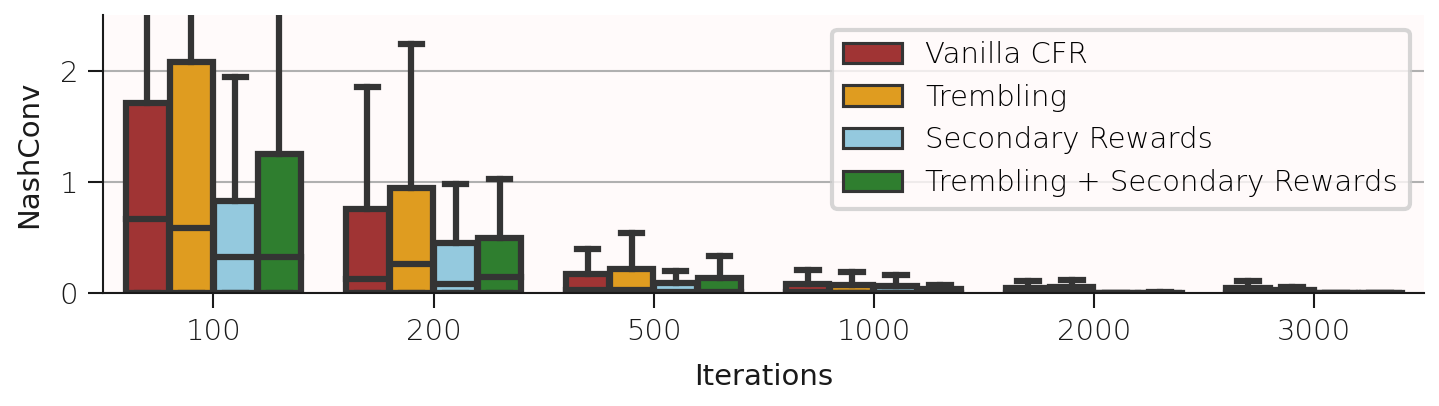}
    \caption{NashConv of MCCFR ablations, varying secondary rewards and trembling opponents.}
    \label{fig:tuning_cfr}
\end{figure}

The results are shown in Figure~\ref{fig:tuning_cfr}.
All four variants consistently converged to approximate equilibria.
However, there were consistent differences between some of the variants: training with secondary rewards led to lower NashConv early in training, and more frequently converged to exact Nash equilibria.
Training against trembling opponents had a more varied impact on convergence, increasing NashConv early in training but having no effect at later checkpoints.
In the remainder of our experiments, we opted to run MCCFR with both of these features, reasoning that they were likely to steer MCCFR away from brittle equilibria and lead to faster convergence.

\subsubsection{PPO: Architecture and Hyperparameter Optimization}
Compared to CFR, PPO has many hyperparameters---controlling its neural network architecture, loss function, and optimizers---and its performance is highly sensitive to these hyperparameter settings.
We tuned these hyperparameters using random search.
Specifically, we sampled 60 random configurations of these hyperparameters and ran each with 3 random seeds on 6 different 5-type games (comprising of 3 valuation samples and 2 bid processing rules), making for a total of 18 runs for each configuration.
For each, we ran PPO for 30 minutes, computing NashConv once at the end.
We provide details about the hyperparameters and our random sampling distributions in Appendix~\ref{sec:ppo_hyperparams}, including more extensive results.

The hyperparameter configurations varied substantially in their ability to find approximate equilibria, with many frequently producing policies with very high NashConv.
However, a few configurations consistently converged to approximate equilibria; the best configuration always returned policies with a NashConv below 0.1.
Notably, this configuration selected the AuctionNet architecture (as do 8 of the best 10 configurations).
We used this single best PPO configuration in the remainder of our experiments.

\subsection{Effects of Bid Processing Algorithms}
\label{subsec:mainresults}

We now turn to our main economic question: \textit{how does the auction's bid processing algorithm affect bidders' strategies and outcomes of the auction?}
To answer this question, we experimented with bidder valuations with 1, 2, 3, 5, and 7 types, generating 5 samples of each for a total of 25 value profiles.\footnote{Bidders' valuations were sampled independently for each of these games: e.g., bidders' values in the 1-type games were not necessarily equal to any values in games with more than 1 type.}
For each value profile, we created one game for each bid processing algorithm.
These games ranged in size from 10--100 information states for the 1-type games to 500--700 for the 7-type games.
We ran MCCFR and PPO on each game with 10 random seeds, for a total of 1000 MARL algorithm runs.
We discarded 24 runs with a NashConv greater than 0.1, leaving 976 runs; 904 of these had a NashConv of 0, meaning that they converged to Nash equilibria.

\begin{figure}
    \centering
    \includegraphics[]{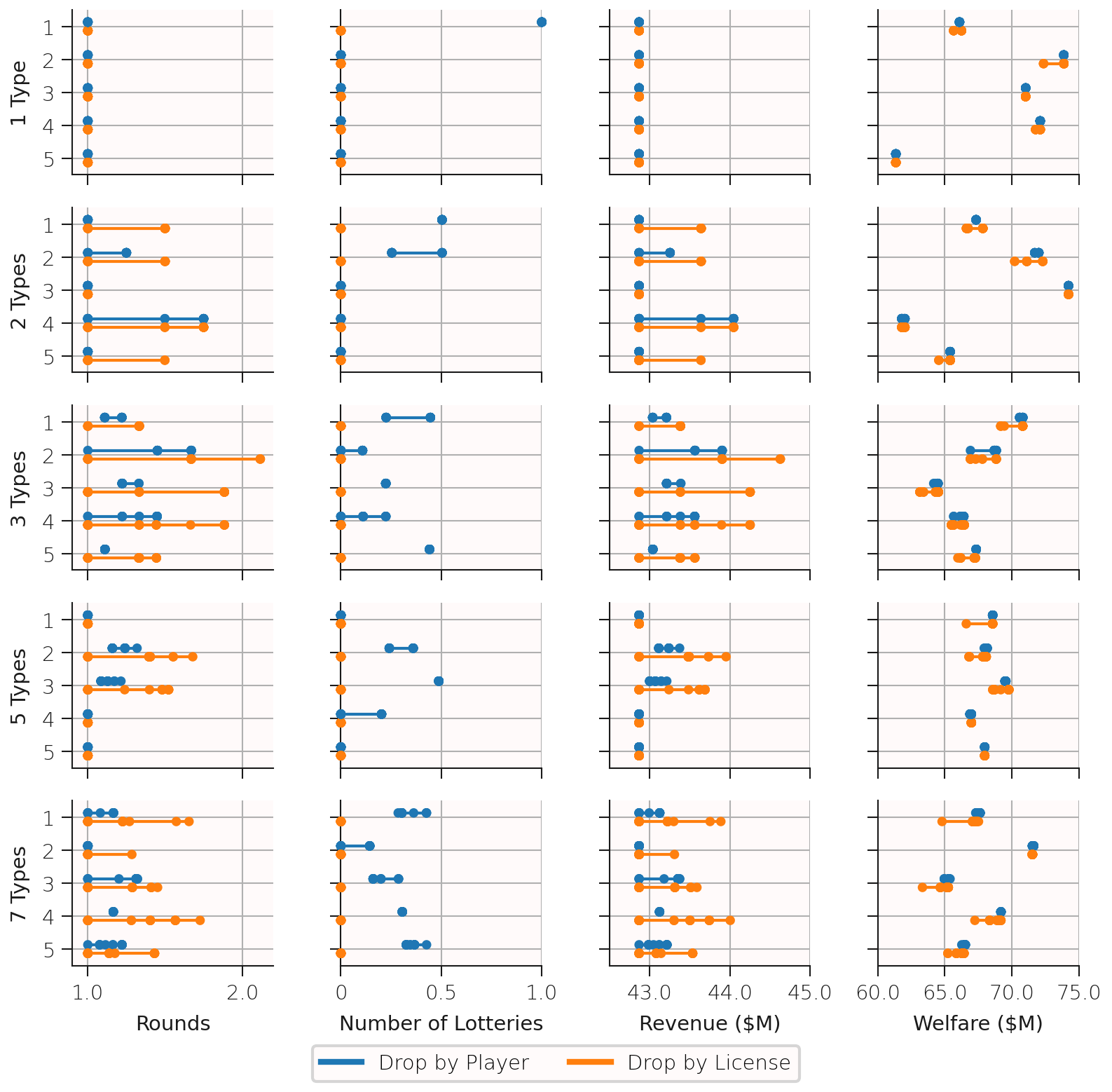}
    \caption{Auction outcomes using MCCFR and PPO on 2-player games with 1 to 7 types.}
    \label{fig:comparative_statics}
\end{figure}

Our results are shown in Figure~\ref{fig:comparative_statics}.
This plot shows four metrics: the length of the auction, number of lotteries (i.e., the number of rounds where the bid processing algorithm can produce two or more different outcomes), revenue, and welfare. These metrics were averaged over 10,000 episodes to account for randomness in bidders' types and bid processing outcomes.
Each point represents an (approximate) equilibrium found by one of the two MARL algorithms. 
We provide supplementary plots showing the equilibria found by each algorithm in Appendix~\ref{sec:extra_results}.

Our main finding is that the two bid processing rules lead to qualitatively different outcomes:
when bidders have 3 or more types, equilibria under the drop-by-license rule can produce longer auctions with higher revenue and lower welfare than under the drop-by-player rule.
These differences in auction outcomes are caused by differences in bidder behavior.
Under drop-by-player, when both bidders attempt to switch their demand from the encumbered product to the unencumbered one, the bid processing algorithm results in a lottery; when this happens, the auction immediately ends.
Under drop-by-license, however, these lotteries are less desirable to bidders, as they risk dropping a single encumbered license which will not create enough activity to buy an unencumbered license.
Bidders adjust their strategies accordingly, taking care to avoid simultaneously switching their demand, causing every single equilibrium under this rule to have no lotteries.
In many games, this adjustment can lead to longer auctions, creating additional revenue as the prices increase.

Notably, the differences between the two bid processing algorithms vanish when bidders only have a single type.
In these games, bidders have complete information about their opponent's values (and, in equilibrium, their opponent's strategy).
They are then able to avoid lotteries regardless of the bid processing algorithm in most games, leading to identical auction lengths and revenues under the two rules.
The results are also ambiguous with 2 types per bidder, where no equilibrium leads to a bid processing lottery in 3 of our 5 games.
These findings highlight the importance of modelling the environment as a Bayesian game with multiple types for each player, as doing so can lead to qualitatively different findings.
Lastly, these findings also highlight the importance of searching for multiple equilibria; if we had only sampled a single equilibrium from each game, we would have drawn very different conclusions.

\begin{figure}
    \centering
    \includegraphics[]{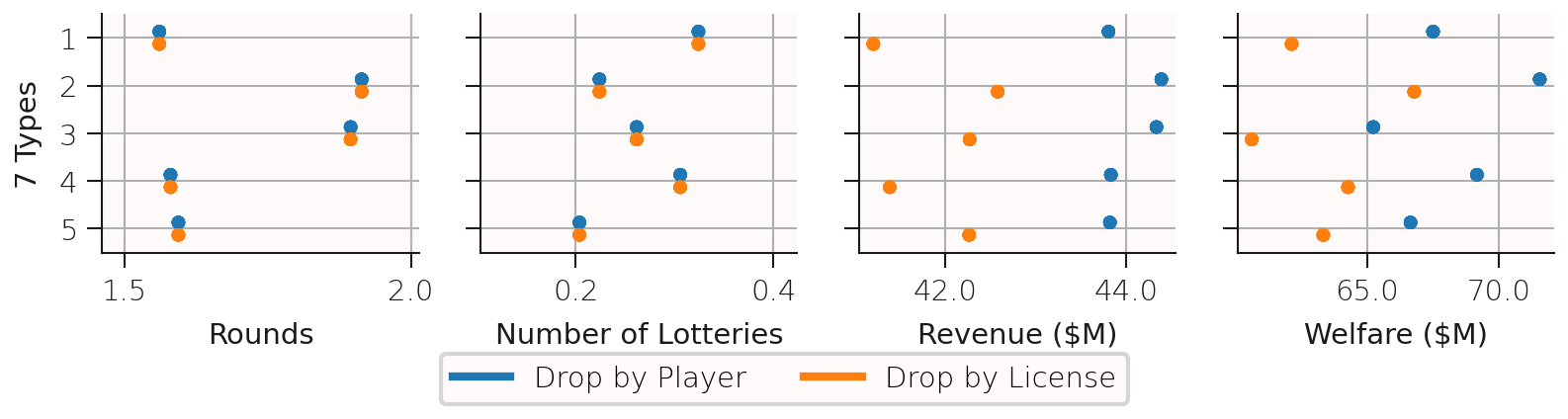}\hfill
    \caption{Auction outcomes under straightforward bidding on 2-player games with 7 types.}
    \label{fig:straightforward}
\end{figure}

We also simulated the auctions under straightforward bidding.
Straightforward simulation results for 7-type games are shown in Figure~\ref{fig:straightforward}, with additional plots for the other games in Appendix~\ref{sec:extra_results}.
Figure~\ref{fig:straightforward} shows that ignoring bidders' strategic incentives would lead to entirely different conclusions.
Here, both rules lead to auctions of the same length, as bidders act the same way regardless of the bid processing algorithm.
Furthermore, it is now the drop-by-player rule that leads to more revenue: under the drop-by-license rule, it is common for a license to be left unsold, losing revenue.
The drop-by-player rule also produces higher welfare for a similar reason, as it is inefficient to leave licenses unsold. 
We note that straightforward bidding is indeed not an equilibrium in these auctions: on the 7-type games, NashConv ranges from 0.18 to 3.10.

\subsection{Scaling to Larger Games}
\label{sec:scale}

Encouraged by the results above, we created a set of larger games to test how well our MARL algorithms would scale.
To do so, we added a third bidder and a fifth encumbered license. 
To keep the games tractable, we also increased the clock increment from 5\% to 20\%.
We initialized the auction in a similar state as before, having each bidder bid for three encumbered licenses for two rounds.
We generated 5 value profiles with 3 types per bidder, yielding a total of 10 games across the two bid processing algorithms.
These games are substantially larger than the 2-player games, ranging from 8,000--40,000 information states.
We ran MCCFR and PPO for 8 hours on each game.

\begin{figure}
    \centering
    \begin{tabular}{cc}
        \includegraphics[scale=0.9]{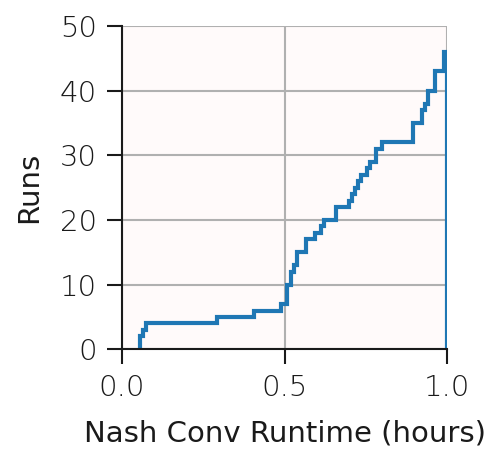} & 
        \includegraphics{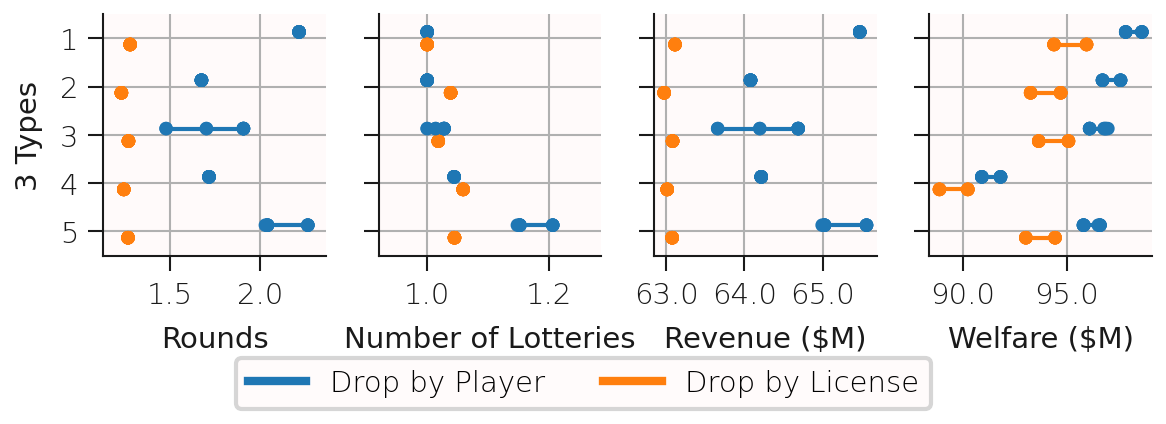} \\
        (a) & 
        (b) 
    \end{tabular}
    \caption{(a) NashConv runtime distribution; (b) auction outcomes, using MCCFR on 3-player games.}
    \label{fig:3p_cfr}
\end{figure}

Validating trained policies on these larger games was substantially more difficult.
At the end of each run, we attempted to compute NashConv for up to 1 hour.
Within this time limit, we were only able to compute NashConv for 46 of the 100 MCCFR runs (Figure~\ref{fig:3p_cfr}a).
When we were able to compute NashConv, we found that MCCFR was consistently able to find (approximate) Nash equilibria: 36 runs had a NashConv of 0, and the highest computed NashConv was 0.03.
Figure~\ref{fig:3p_cfr}b shows the auction metrics for all 100 MCCFR runs, including those for which we could not compute NashConv.
Notably, the differences between the rules in this setting are quite unlike those in the 2-player games, with the drop-by-player rule leading to longer auctions, higher revenue, and higher welfare.
One potential explanation is that under the drop-by-player rule, all three players can safely attempt to switch their demand from the encumbered product to the unencumbered product, allowing for further competition in the following round.

We found that PPO performed far worse than MCCFR.
We were able to compute NashConv for 37 of 50 PPO runs, finding that it was unable to reliably find approximate equilibria: almost all runs had a NashConv greater than 3, and the majority had a NashConv greater than 10.
Thus, it appears nontrivial to scale PPO to larger games; we suspect that its hyperparameters would need to be retuned for this larger environment in order to perform well.

\section{Conclusions and Discussion}\label{sec:rl_discussion}

Multiagent reinforcement learning algorithms are rapidly maturing, having recently reached several significant milestones. 
This paper argues that---with the right modeling and algorithmic choices---MARL can be used to analyze iterative combinatorial auctions that resist both pen-and-paper and computational Bayes--Nash analysis. 
We laid out a methodology for modelling an auction for use with MARL analysis, finding near-equilibria using MARL algorithms, and both validating and interpreting the results. 
We then offered a case study on bid processing in clock auctions, showing that MARL analysis could reach an economically meaningful conclusion in a setting that does not admit analysis with existing approaches, and for which a simulation-based analysis would have yielded the wrong conclusion, with potential to scale to even larger settings.

We hope that we have persuaded the reader that MARL has something to offer auction analysis; however, our own work has clearly only scratched the surface. We thus conclude with a discussion of useful next steps: future clock auction design choices that would be fruitful to analyze; promising avenues for improving our methodology; and open theoretical questions.

\subsection{Promising Auction Design Questions}
As we described in Section~\ref{sec:rl_clock}, an auction designer faces many potential choices about how to instantiate each aspect of an auction, each of which could influence bidding behavior and ultimately outcomes. 
Here, we describe three such design choices that we believe would be amenable to MARL analysis, demonstrating the relevance of each with citations either to the academic literature or to bidder consultations with policymakers.
For each, we discuss the design choice's strategic implications and how it might impact revenue, welfare, and/or length of auctions. 

\paragraph{Clock speed} 
A designer chooses how quickly the prices of overdemanded products increase. 
Larger increments allow for shorter auctions, but give less feedback to bidders, reducing opportunities for price discovery. 
Bidders react to these increments: in a consultation for the 3800 MHz Canadian auction, Telus ``argued that increments [of 10\%-20\%] would create an accelerated cadence at the start of the auction that may prove contrary to the desired intention of promoting price discovery'' \cite{ised3800}. 
An auctioneer may even consider dynamic schemes, modifying the clock speed throughout the auction, or using steeper price increases for more overdemanded products. 

\paragraph{Information disclosure} 
At the end of each round, the auctioneer discloses information about the current state of the auction. 
Commonly, this includes the prices and units of aggregate demand for each product. 
An auctioneer could vary this policy---for example, only disclosing whether each product was over- or under-demanded---hoping to improve outcomes by disincentivizing strategic bidding.
An opposite extreme is to deanonymize all bids. 
Such a suggestion was made in a consultation for the 3500 MHz Canadian auction: ``Eastlink opposed the use of anonymous bidding, stating that it generally favours the large national incumbent bidders as they inherently have more information available to them during the price discovery rounds'' \cite{ised3500}. 

\paragraph{Activity rules} 
There is a body of work on how to select activity rules, mostly centered around which one to pick from the perspectives of discouraging strategic manipulations and minimizing complexity for the bidder. 
For example, \citet{parkesactivity} define \emph{strong activity rules} as those which admit straightforward bidding strategies and exclude strategies that not consistent with straightforward bidding under some utility function. 
\citet{garpunique} propose that an activity rule should (1) ensure demand and price move in opposite directions, (2) always allow straightforward bidding, and (3) always allow repeating the previous bid; the authors prove that an activity rule based on the Generalized Axiom of Revealed Preference (GARP) uniquely satisfies the three axioms under non-linear prices. Of course, neither of these papers is backed by equilibrium analysis, as this would be intractable; it would thus be valuable to use MARL analysis to estimate the extent to which these activity rules allow strategic manipulation.

It would also be valuable to study bidder behavior under a variety of value models---for example, modelling risk-averse bidders, or spiteful bidders who gain utility from competitors paying high prices---to understand the extent to which any findings are robust to the choice of bidder model.

\subsection{Methodological Improvements}
\subsubsection{Scaling to Larger Games.}
A limitation of our experiments is their scale: 
even our larger games (Section~\ref{sec:scale}) only had tens of thousands of information states, relatively small by reinforcement learning's standards.
However, scaling to larger games presents many challenges.
As the game tree grows large, tabular methods require more space to store their policies;
it becomes infeasible complete even a single iteration of a tree-traversal method;
and computing NashConv exactly becomes unrealistic, complicating both analysis and hyperparameter tuning.
This final point is especially impactful for deep learning methods, which typically need to have their hyperparameters re-tuned for different environments.
Future work should study the extent to which these challenges can be overcome, scaling MARL algorithms to larger and more complex auctions.

\subsubsection{Enumerating many equilibria.} \label{subsec:quality_diversity}
We found multiple equilibria by running MARL algorithms with different random seeds. 
This approach is easy to implement, but may not be able to find all equilibria of a game.
Furthermore, inductive biases in the algorithm make it more likely to find some equilibria than others.
While one might believe that equilibria which are hard to find are less plausible, this may also be seen as a blind spot.
Future work should develop and evaluate algorithms to increase the diversity of equilibria, expanding our ability to fully characterize the equilibria of an auction.
For example, it may be fruitful to adapt ideas from \emph{quality diversity} in single-agent RL~\cite[e.g.,][]{pugh2016quality}, which aims to find multiple distinct, good policies.

\subsubsection{Interpreting bidding behavior.}
While our case study focused mostly on auction outcomes, it can also be important to derive insights about what constitutes a good strategy. 
This requires that learned policies be not only measurable, but interpretable. 
\citet{bertsimasworld} fit decision trees to trained CFR strategies to produce interpretable poker strategies. 
Future work could similarly focus on distilling bidding behavior. 
A set of composable building blocks of bidding strategies could be valuable both for understanding auctions and to inform how to shrink bidders' action space in larger-scale experiments.

\subsection{Theoretical Questions}
\subsubsection{Convergence to Nash equilibrium.}
In two-player zero-sum games, CFR approaches a Nash equilibrium.
However, auctions are not zero-sum, and existing guarantees for general-sum games are far weaker. 
In this case, \citet{morrill2021hindsight} show that CFR is only guaranteed to converge to a counterfactual coarse correlated equilibrum, a much weaker solution concept.
Despite this lack of theoretical justification, on our auctions, we found that MCCFR consistently converged to pure Bayes--Nash (near-)equilibrium.
Is there a class of games (larger than two-player zero-sum) in which existing MARL algorithms converge to Nash equilibria? 
Understanding the space of such games would help with selecting an appropriate MARL algorithm.

\subsubsection{Alternative solution concepts.}
While we have focused on finding (Bayes--)Nash equilibria, this solution concept can admit unrealistic behavior.
One concrete problem is that it is possible for our learned policies to make non-credible threats in off-path states.
For example, a bidder could adopt a trigger strategy, bidding round after round to increase prices to unprofitable heights if their opponent does not coordinate with them in earlier rounds.
These strategies are unrealistic, as no reasonable bidder would go bankrupt to punish an opponent.
Various equilibrium refinements seek to eliminate this behavior, such as sequential equilibrium~\cite{sequentialeq}; developing MARL algorithms that converge to these refinements would be a valuable direction for future work.

%
\begin{acks}
Thanks to Marc Lanctot and Betty Shea for their technical contributions on early iterations of this work, and to Costa Huang, Paul Milgrom, Julien Perolat, Samuel Sokota, Andrew Vogt, and Albert Zuo for helpful discussions.
This work was funded by an NSERC CGS-D scholarship, an NSERC Discovery Grant, a DND/NSERC Discovery Grant Supplement, a CIFAR Canada AI Research Chair (Alberta Machine Intelligence Institute), a Digital Research Alliance of Canada RAC Allocation, awards from Facebook Research and Amazon Research, DARPA award FA8750-19-2-0222, CFDA \#12.910 (Air Force Research Laboratory), and through computational resources and services provided by Advanced Research Computing at the University of British Columbia.
This work was done in part while several authors were visiting the Simons Institute for the Theory of Computing.
\end{acks}

\bibliographystyle{ACM-Reference-Format}
\bibliography{cfr}

\appendix

\section{Computational Environment}\label{sec:computationalenv}
Our experiments were performed on two different compute clusters. 
Our first cluster consisted of nodes equipped with 32 2.10GHz Intel Xeon E5-2683 v4 CPUs with 40960 KB cache and 96 GB RAM.
Our second cluster consisted of 16-core machines with Intel Silver 4216 Cascade Lake 2.1GHz processors and 96 GB RAM. 
Each run had access to 4 CPUs and 20 GB of RAM.

\section{Clock Auctions}\label{sec:clockauctions}

In this appendix, we give a more formal description of the clock auction games that we implemented.

Let $N$ be the number of bidders and $M$ be the number of service areas.
Each service area $j$ contains $q_j$ identical licenses for sale.\footnote{A realistic auction, e.g. Canada's 600 MHz auction had $N=~10$ and $M=~15$. Canada's 3500 MHz auction had $M$ much higher, in the hundreds). Typical values for $q_j$ range from 1 to 10.} 
The clock auction proceeds over a series of rounds $t$, beginning with $t=1$. 
In each round $t$, all bidders observe a \textit{start-of-round price} $P_{j,t}$ and a \textit{clock price} $P_{j,t} \cdot a$ for each product $j$, where $a$ is the \textit{clock increment} (e.g., 5--20\%).
(These prices are anonymous---i.e., all bidders face the same price.)
Then, each bidder $i$ simultaneously submits a bid $B_{i,t,j} \in \{0, 1, \dots, q_j\}$ for each product $j$, representing their demand for each product.
After the bids are made, each product has an aggregate demand $Z_{j,t} = \sum_i B_{i,j,t}$.
If no product is overdemanded (i.e., $Z_{j,t} \le q_j$ for all products $j$), then the auction ends, with each bidder $i$ winning $B_{i,t,j}$ licenses of product $j$, paying $P_{t,j}$ for each one.
Otherwise, the price of each overdemanded product increases, with the start-of-round price being set to the previous clock price, and the auction continues on to the next round.

Typically, auctions include an \textit{activity} rule. 
The rule restricts the set of legal bids based on a bidder’s bid history with the aim of discouraging strategic bidding.
While there are many variants, a simple and common rule is to assign each product $j$ a number of \textit{eligibility points} $e_j$; then, each bidder's total activity $\sum_j B_{i,t,j} \cdot e_j$ must be non-increasing. 
This rule disallows bidders from bidding for few items early in the auction and rapidly expanding their bid later. 

Auctions can allow for intra-round bidding. Such a provision allows bidders to express changes in their demand between clock increments (e.g., if the start-of-round price is \$100 and the clock price is \$120 and a bidder holds two licenses in a region, they may bid to drop to one license at \$110. If this drop causes supply to equal demand, the new start-of-round price would be \$110.). Intra-round bidding allows an auctioneer to set larger clock increments. Perhaps our most significant modeling simplification is that we do not model intra-round bidding. Instead, we assume that agents can \textit{only} bid at the clock price. This restriction requires the clock increments in our modeled games to be kept relatively small.

The clock auction includes a rule that prohibits demand from dropping ever dropping from above supply to below supply. This guarantees that once a product has ever had a round where demand was at least supply, it will be sold. In practice, this requires the auctioneer to reject some bids. The bid processing algorithm works by placing tuples of \textit{requests} of the form (bidder, product, change in demand) into a queue. These requests are then processed in queue order to the degree possible (a request may only be partially fulfilled if either there is not enough demand to allow further drop bids, or the bidder does not have sufficient activity to allow a pick up bid). If a request is only partially fulfilled, it is reinserted into the queue. The algorithm continues applying these requests until it is not able to apply any additional requests in the queue. 

\section{Value Sampling Details}
\label{sec:sats_details}

\paragraph{Removing uninteresting games.}
Our game sampler rejects games that we deem strategically uninteresting. 
We require that every bidder, in all type realizations, is allocated at least one license under the utility-maximizing allocation at opening prices. 
We solve for the utility-maximizing allocation using a MIP. 
A bidder that wins nothing is hopelessly weak, slowing down simulations for no added benefit.

\paragraph{Infeasible games.}
Some criteria ensured that we generated games that we were capable of solving. 
First, as a proxy measure for the length of the auction, we computed the number of rounds it would take for the auction to conclude if each bidder bid for its profit-maximizing bundle at start-of-round prices in every round.
We rejected auctions longer than 20 such rounds.
Second, we reran the above simulation, only incrementing the price on a single product randomly chosen among those that were overdemanded in each round. 
This is a crude model of how long the auction could last in the worst case.
We rejected simulations that took on average longer than 25 rounds to conclude.
Lastly, for 2-player games, we required MCCFR to be able to run at least 25 iterations per second.
These checks were all based on the drop-by-player rules.

\paragraph{Changes for 3-player games.}
We made several small changes to the value model for the 3-player games in Section~\ref{sec:scale}.
For MRVM parameters, we sampled market shares uniformly from 20\% to 30\% and values per subscriber uniformly from \$35 million to \$45 million.
The keypoints on their sigmoidal value function were placed at $\pm$10\% from their market share.
We only required MCCFR run at 1 iteration per second rather than 25. 

\begin{figure}
    \centering
    \includegraphics[width=8cm]{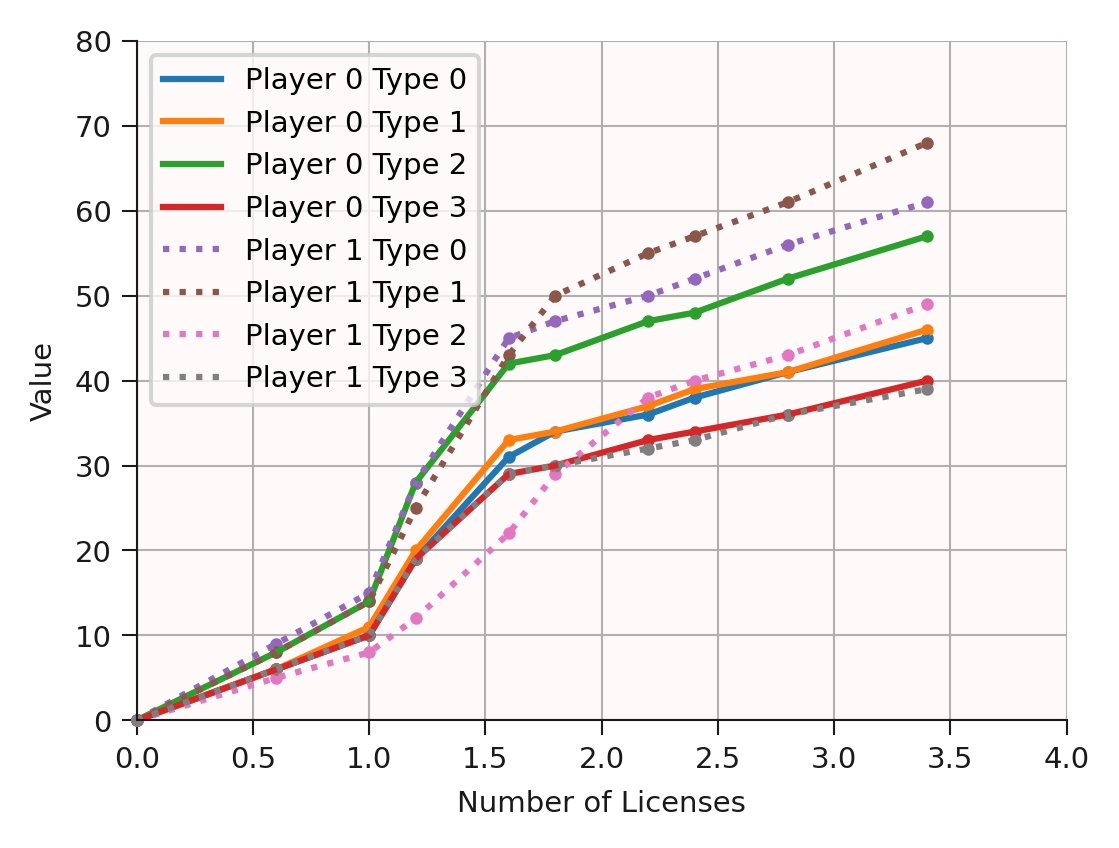}\hfill
    \caption{An example value profile for a game with two bidders and four types. Fractional numbers of licenses on the x-axis correspond to bundles with encumbered licenses.}\label{fig:values}
\end{figure}

\paragraph{Example value function.}
We show an sample value function in Figure~\ref{fig:values}. 

\section{PPO Features}
\label{sec:ppo_features}
Table~\ref{tab:ppo_features} lists the features describing an information state that were provided to each PPO agent.
All features were normalized such that typical values ranged from 0 to 1.

\begin{table}[H]
    \centering
    \begin{tabular}{lll}
        \toprule
        \textbf{Feature} & \textbf{Description} & \textbf{\# Variables} \\
        \midrule
        Bundle Quantities & Number of licenses included in each bundle. & $P \times B$ \\
        Bundle Activity & Total activity points of each bundle. & $B$ \\
        Values & Bidder's value for each bundle. & $B$ \\

        Prices & History of prices for each bundle. & $T \times B$ \\
        Submitted Demands & This bidder's bid in each round, one-hot encoded. & $T \times B$ \\ 
        Processed Demands & This bidder's bid in each round, after applying bid processing. & $T \times B$ \\
        Aggregate Demands & Sum of all bidders' processed demands in each round. & $T \times P$ \\ 
        
        Current Activity & Total activity points of this bidder's last processed bundle. & 1 \\
        Current Exposure & Start-of-round cost of this bidder's last processed bundle. & $1$ \\

        \bottomrule
    \end{tabular}
    \caption{Features provided to each PPO agent. In the \# Variables column, $P$ denotes the number of distinct products; $B$ denotes the total number of bundles; and $T$ denotes the maximum length of the auction.}
    \label{tab:ppo_features}
\end{table}

\section{PPO Hyperparameters}
\label{sec:ppo_hyperparams}

Our configurations were sampled as follows:
\begin{lstlisting}[basicstyle=\footnotesize]
config['steps_per_batch'] = int(np.random.choice([64, 128]))
config['num_minibatches'] = int(np.random.choice([4, 8]))
config['update_epochs'] = int(np.random.choice([4, 8, 16]))
config['learning_rate'] = float(np.random.choice([3e-5, 6e-6, 9e-5, 1e-4, 3e-4]))
config['gae'] = bool(np.random.choice([True, False]))
config['anneal_lr'] = bool(np.random.choice([True, False]))
config['gae_lambda'] = float(np.random.uniform(0.94, 1.))
config['clip_coef'] = float(loguniform.rvs(0.0003, .3, size=1)[0])
config['clip_vloss'] = bool(np.random.choice([True, False]))
config['entropy_coef'] = float(np.random.choice([1e-4, 1e-5, 1e-6, 0, 3e-5, 3e-6]))
config['value_coef'] = float(loguniform.rvs(0.1, 1.3, size=1)[0])
config['num_envs'] = int(np.random.choice([4, 8, 16]))
config['normalize_advantages'] = bool(np.random.choice([True, False]))

config['optimizer'] = random.choice(['adam', 'rmsprop', 'sgd'])
if config['optimizer'] == 'adam':
    beta1 = float(np.random.uniform(0.8, .9))
    beta2 = float(np.random.uniform(0.8, .999))
    config['optimizer_kwargs'] = {'betas': [beta1, beta2]}
config['max_grad_norm'] = float(np.random.uniform(0.1, 1)) 

config['agent_fn_kwargs'] = dict()
agent_fn = base_config['agent_fn'].lower()
if agent_fn == 'auctionnet':
    config['agent_fn_kwargs']['activation'] = str(np.random.choice(
        ['relu', 'tanh']
    ))
    config['agent_fn_kwargs']['hidden_sizes'] = random.choice(
        [[32, 32, 32], [64, 64, 64], [128, 128, 128]
    ])
    config['agent_fn_kwargs']['add_skip_connections'] = bool(np.random.choice(
        [True, False]
    ))
    config['agent_fn_kwargs']['use_torso'] = bool(np.random.choice([True, False]))
elif agent_fn == 'ppoagent': # MLP
    hidden_sizes = random.choice([
        [64, 64], [128, 128], [256, 256], 
        [64, 64, 64], [128, 128, 128], [256, 256, 256],
    ])
    activation = str(np.random.choice(['relu', 'tanh']))
    config['agent_fn_kwargs']['actor_hidden_sizes'] = hidden_sizes
    config['agent_fn_kwargs']['critic_hidden_sizes'] = hidden_sizes
    config['agent_fn_kwargs']['actor_activation'] = activation
    config['agent_fn_kwargs']['critic_activation'] = activation

\end{lstlisting}

\begin{figure}
    \centering
    \includegraphics[width=\textwidth]{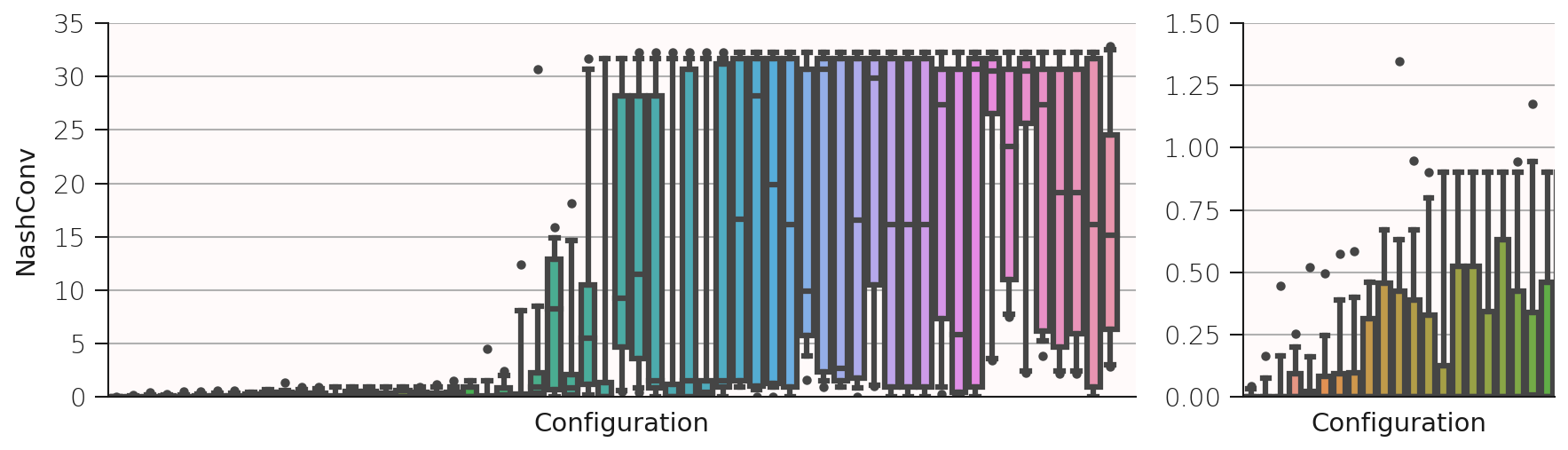}
    \caption{NashConv of PPO hyperparameter tuning runs, showing all 60 configurations (left) and a detailed view of the 20 best configurations (right). Configurations are sorted by 95th percentile of NashConv. Whiskers denote 5th and 95th percentiles.}
    \label{fig:tuning_ppo}
\end{figure}

Our final PPO configuration is described in Table~\ref{tab:ppo_hyperparams}, and the performance of 60 randomly sampled configurations is plotted in Figure~\ref{fig:tuning_ppo}.

\begin{table}
    \begin{tabular}{ll}
        \toprule
            Variable & Value \\
        \midrule
            Activation & relu \\
            Add Skip Connections & True \\
            Anneal LR & True \\
            Architecture & AuctionNet \\
            Clip Vloss & False \\
            Entropy Coef & $3.0 \times 10^{-5}$ \\
            Gae & False \\
            Gamma & 1.0 \\
            Hidden Sizes & [32, 32, 32] \\
            Learning Rate & $9 \times 10^{-5}$ \\
            Max Grad Norm & 0.526 \\
            Normalize Advantages & False \\
            Num Envs & 16 \\
            Num Minibatches & 8 \\
            Optimizer & adam \\
            Optimizer Betas & [0.803, 0.865] \\
            Steps Per Batch & 64 \\
            Track Stats & True \\
            Update Epochs & 8 \\
            Use Torso & True \\
            Value Coef & 0.609 \\
        \bottomrule
    \end{tabular}
    \caption{Our best-performing hyperparameters for PPO.}        
    \label{tab:ppo_hyperparams}
\end{table}

\section{Additional Results}
\label{sec:extra_results}
\begin{figure}
    \centering
    \includegraphics[]{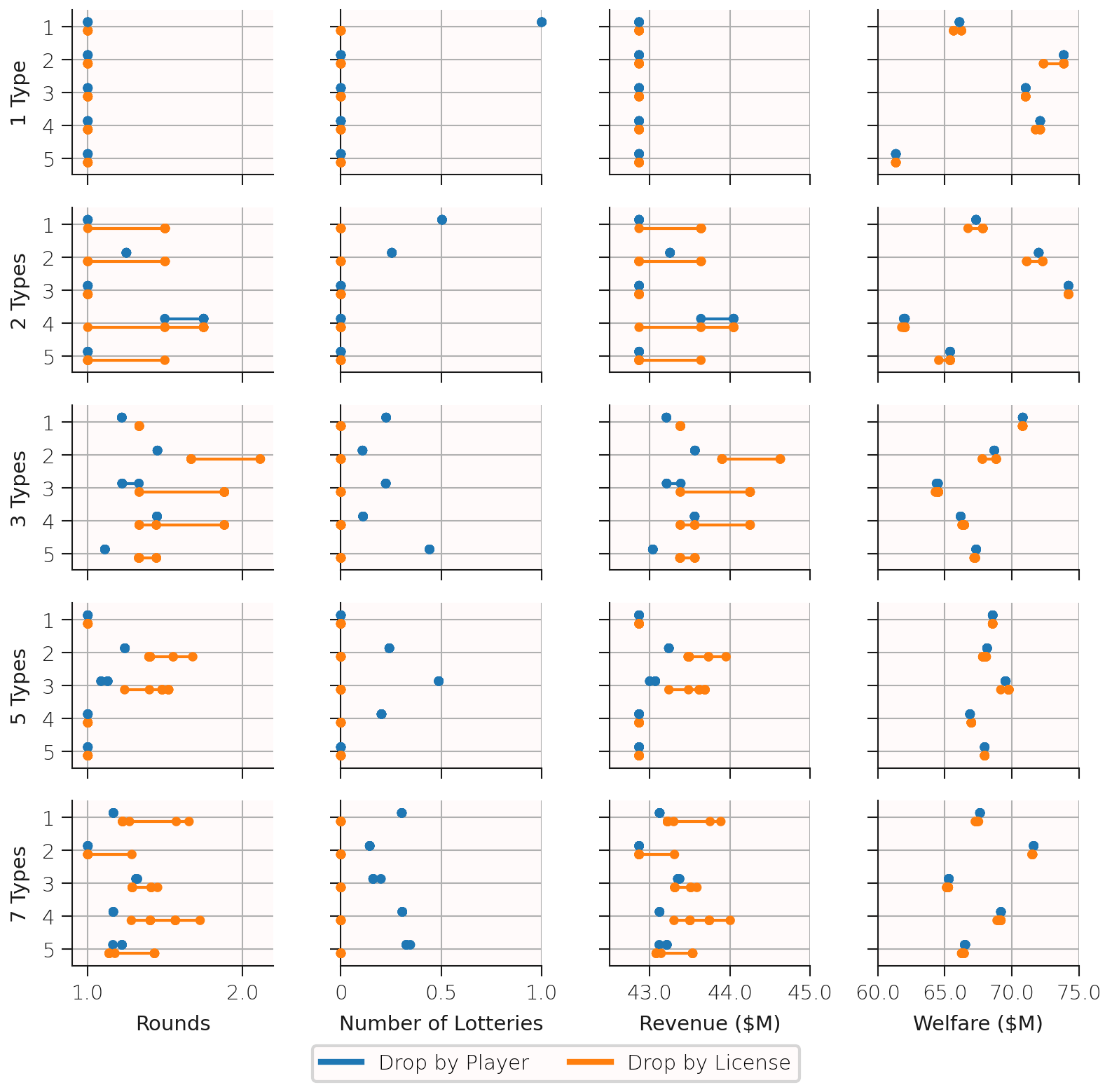}
    \caption{Auction outcomes using MCCFR on 2-player games with 1 to 7 types.}
    \label{fig:comparative_statics_cfr}
\end{figure}

\begin{figure}
    \centering
    \includegraphics[]{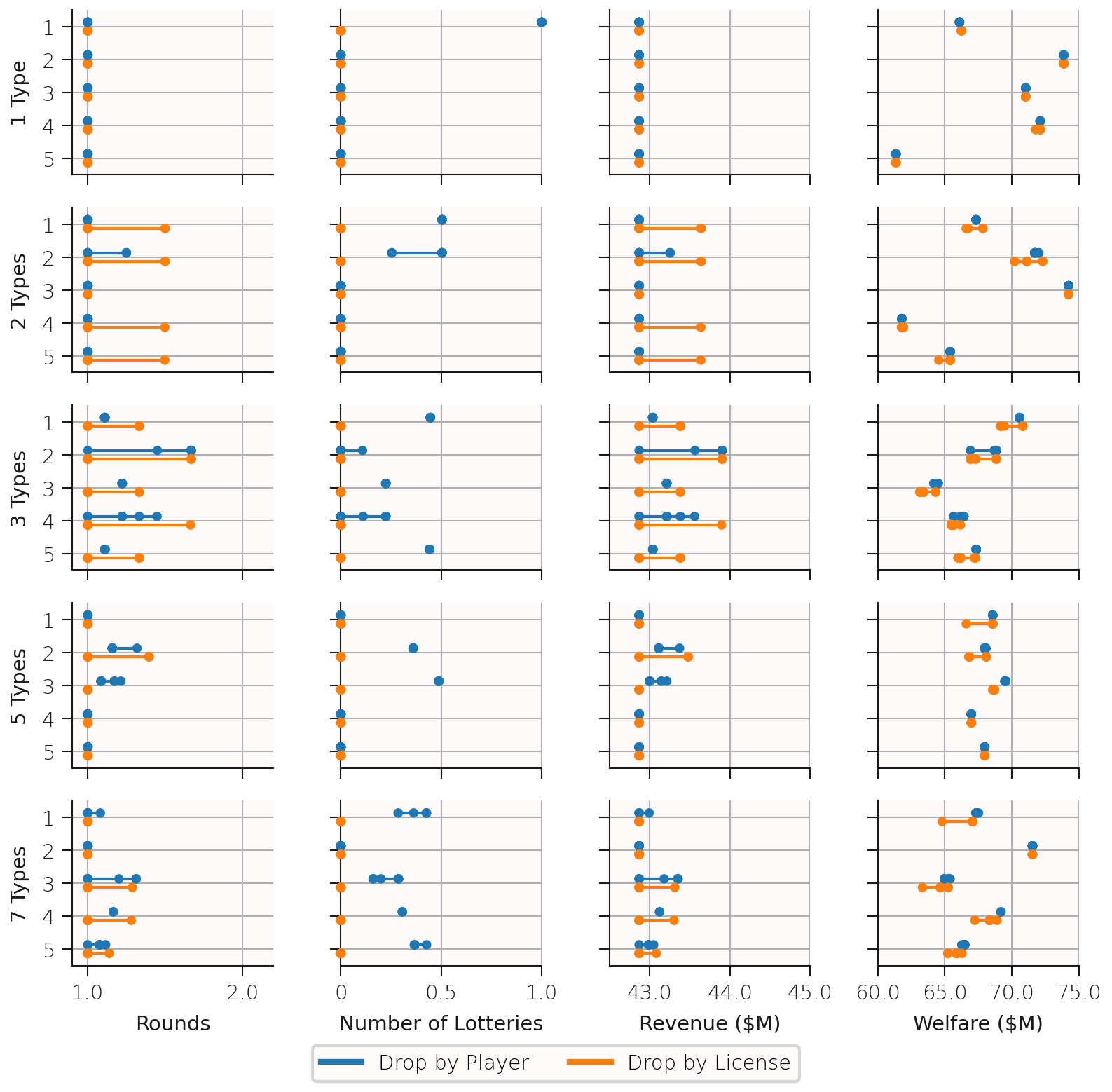}
    \caption{Auction outcomes using PPO on 2-player games with 1 to 7 types.}
    \label{fig:comparative_statics_ppo}
\end{figure}

\begin{figure}
    \centering
    \includegraphics[]{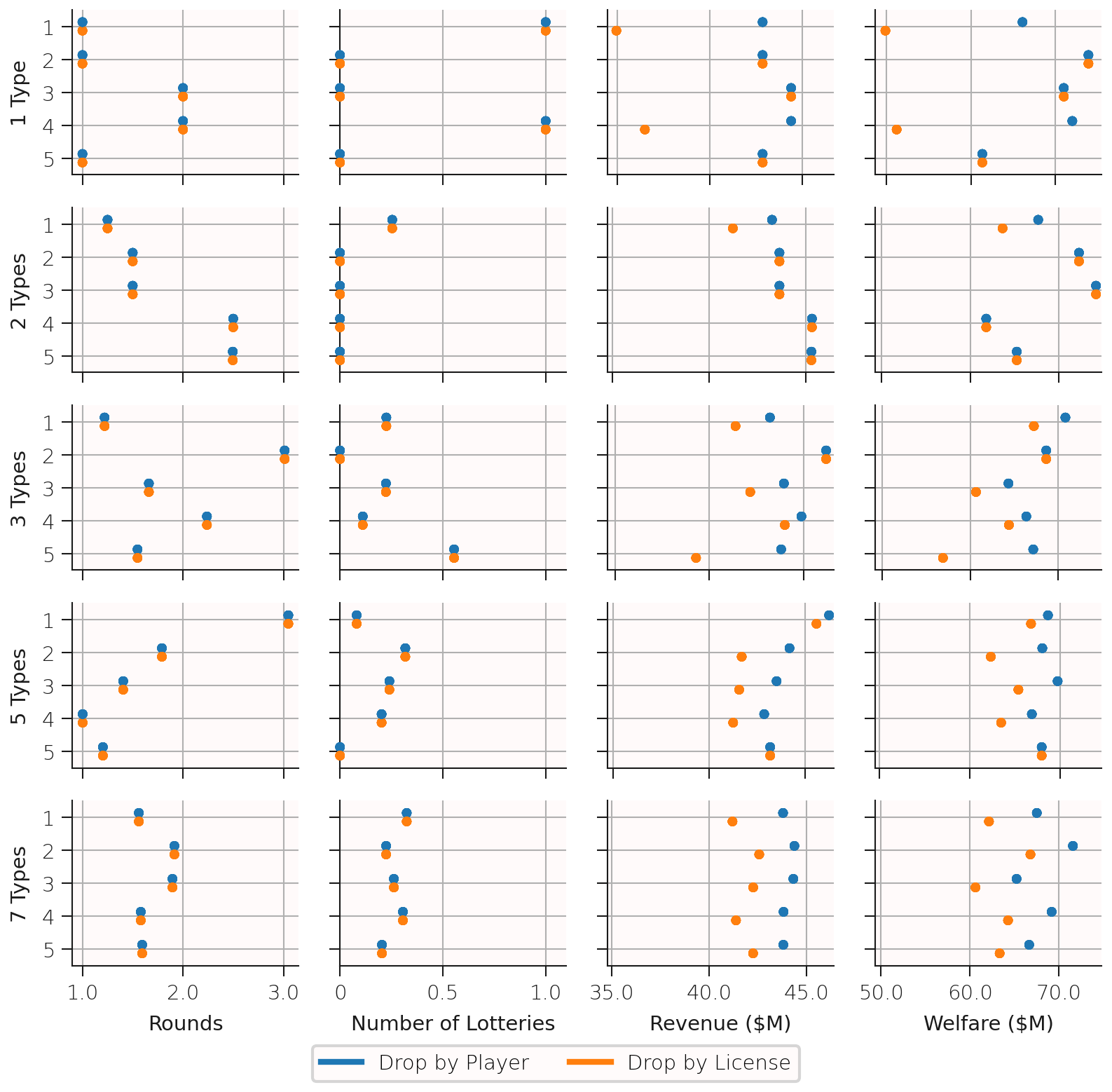}
    \caption{Auction outcomes under straightforward bidding on 2-player games with 1 to 7 types.}
    \label{fig:comparative_statics_straightforward}
\end{figure}

\begin{figure}
    \centering
    \includegraphics[]{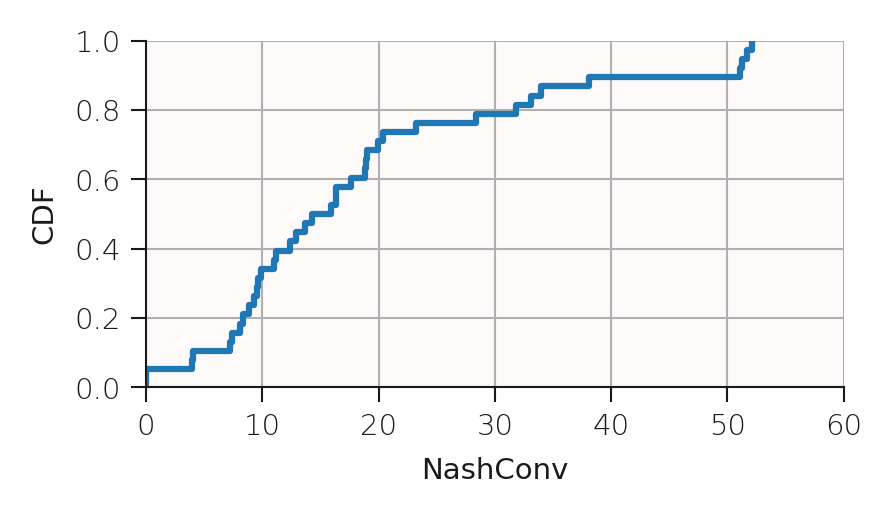}
    \caption{Distribution of NashConv for PPO on 3-player games, including only runs for which NashConv could be computed in under 1 hour.}
    \label{fig:ppo_3p_nashconv}
\end{figure}

In Figures~\ref{fig:comparative_statics_cfr} and~\ref{fig:comparative_statics_ppo}, we show auction outcomes on the 2-player games, separating equilibria found by each of the two MARL algorithms. 
Interestingly, we note that there exist some equilibria that are only found by one of the two algorithms.
Figure~\ref{fig:comparative_statics_straightforward} shows auction outcomes on these games assuming bidders follow a straightforward bidding heuristic.
Finally, Figure~\ref{fig:ppo_3p_nashconv} shows the distribution of NashConv for PPO runs on the 2-player games.

\end{document}